\documentclass[12pt]{article}

\newcommand{\bea}{\begin{eqnarray}}
\newcommand{\eea}{\end{eqnarray}}

\begin{document}

\title{\bf Comment on the Alday-Maldacena solution in
calculating scattering amplitude via AdS/CFT}
\author{Gang Yang\thanks{E-mail: yangg@itp.ac.cn} }
\date{}

\maketitle

\centerline{\it Institute of Theoretical Physics, Chinese Academy of
Sciences} \centerline{\it P. O. Box 2735, Beijing 100080, P. R.
China}

\begin{abstract}
Following the recent proposal of Alday and Maldacena to obtain the
strong coupling scattering amplitude in ${\cal N}$ = 4 SYM via
AdS/CFT, we point out that a unique solution can be obtained by
imposing all the Virasoro constraints. In the case of four-gluon
scattering, this solution is identical to the Alday-Maldacena
solution, which is in accordance with the ansatz of Bern, Dixon and
Smirnov. This also solves the moduli space problem of the four-point
solution in a recent paper of Mironov, Morozov and Tomaras.
\end{abstract}

\bigskip

Recently Alday and Maldacena proposed a novel method to calculate
planar gluon scattering amplitudes at strong coupling in ${\cal N}$
= 4 SYM by using AdS/CFT duality \cite{AM}. At leading order the
calculation is reduced to finding the minimal area of a string with
a light-like boundary.

On the other hand, in ${\cal N}$ = 4 SYM, an ansatz for the all
order form of the $n$ gluon MHV scattering amplitudes has already
been given by Bern, Dixon and Smirnov \cite{BDS}. This ansatz (the
BDS ansatz) was supposed to be valid at both weak and strong
coupling. In the weak coupling regime, it has been verified for the
four-point amplitude up to five-loop and five-point amplitude up to
two-loop \cite{4p2l}--\cite{5p2l}.

By using their proposal \cite{AM}, Alday and Maldacena computed the
explicit form of the amplitude for the scattering of four gluons and
found precise agreement with the BDS ansatz to the leading order of
strong coupling. Inspired by this new correspondence, there have
appeared a number of closely related works and generalizations in
\cite{Abel:2007mw}--\cite{Kluson:2008wv}. In particular, an
important quantity in the BDS ansatz, the one-loop MHV $n$-gluon
amplitude, can be written as a double contour integral along a
polygonal Wilson loop $\Pi$, which is defined by the external gluons
momenta \cite{Brandhuber:2007yx}(see also \cite{Drummond:2007au}):
\bea M^{(1)}_n = \oint_\Pi\oint_\Pi \frac{dy^\mu dy'_\mu}
{[(y-y')^2]^{1+\epsilon}} \label{loopweak} \eea
By the proposal of Alday and Maldacena \cite{AM}, this geometrical
integral should be identified with another geometric quantity: the
minimal area of a string in $AdS_5$ which is bounded by the same
polygon (see \cite{MMT} for more details).

To realize the proposal of Alday and Maldacena, it is essential to
find the classical string solution with given boundary conditions.
In \cite{AM}, the solution of four-gluon scattering was found by
doing conformal transformations to a cusp solution, or by trial and
error. For the general multi-gluon scattering, it's more
difficult to find solutions%
\footnote{There have been discussions on the solutions of the large
n-point case in \cite{AM2}, and 6-point and 8-point case in
\cite{ADIN}. The interesting dressing method for finding new
possible solutions was discussed in \cite{Jevicki}.}.
Due to the lack of a general method to solve the complicated
equations, it is also not clear whether the solution is unique or
not.

In a recent paper \cite{MMT}, Mironov, Morozov and Tomaras solved
the sigma-model equations of motion in the case of four-gluon
scattering by using a special ansatz. Surprisingly, the solution was
found to have a moduli space $\{z_a, \phi\}$ \cite{MMT}, and
moreover, the regularized minimal area is also moduli dependent.
This raises a problem: which solution in the moduli space is the
`right' solution that corresponds to the unique scattering
amplitude? In \cite{MMT}, the authors suggested that the
Alday-Maldacena solution could be considered as a minimum of the
regularized action in the moduli space.

In this paper, we point out that a unique solution can be obtained
by imposing all the  Virasoro constraints%
\footnote{The Virasoro constraints were also considered in
\cite{MMT2} when the relation between the $\sigma$-model action and
Nambu-Goto action was discussed. For the Nambu-Goto action, due to
the reparametrization invariance, we can always choose a
parametrization to give the same equations of motion plus the
Virasoro constraints as that from the string $\sigma$-model action.
}.
We will show this explicitly in the case of four-gluon scattering,
where the moduli space variables $\{z_a, \phi\}$ in \cite{MMT} can
be fixed uniquely by the Virasoro constraints. This is supposed to
be true in the cases of general multi-gluon scattering.

\bigskip
\bigskip

We first give a short review of the solution in \cite{MMT}. We will
follow closely the notations used in \cite{MMT}.

The string $\sigma$-model action is
\bea S[X,g] = \int d^2 u ~ \sqrt{g} ~ g^{ij}~ G_{MN} \partial_i X^M
\partial_j X^N , \eea
where the world-sheet metric $g_{ij}(u_1,u_2)$ is taken to be
Euclidean.

In conformal gauge and for the $AdS_5$ target space, the string
$\sigma$-model action takes the following form
\bea S = \int d^2 u~ \frac{({\vec \partial} r)^2+({\vec \partial}
{\bf y})^2}{r^2} . \eea
The bold font is for 4d vectors in the target space, while arrow is
used for 2d vectors on the world-sheet.

The equations of motion are
\bea & & {\vec \partial} \left(\frac{{\vec \partial} r}{r^2}\right)
= -\frac{L}{r} , \quad {\vec \partial} \left(\frac{{\vec \partial}
{\bf y}}{r^2}\right) = 0 , \\ & & L = {({\vec \partial} r)^2 +
({\vec \partial} {\bf y})^2 \over r^2} . \eea

The solution should also satisfy the Virasoro constraints, i.e.
$\delta_g S[X,g]=0$, which in conformal gauge read
\bea & & (\partial_1 r)^2 - (\partial_2 r)^2 + (\partial_1 {\bf
y})^2 - (\partial_2 {\bf y})^2 = 0 , \\ & & \partial_1 r \partial_2
r + \partial_1 {\bf y} \partial_2 {\bf y} = 0 . \eea

In coordinate $z = {1/r}, {\bf v} = {{\bf y}/ r}$, the equations of
motion take the following form
\bea & & \Delta z = zL , \quad \Delta {\bf v} = {\bf v} L \label{eomzv} , \\
& & z^2L-({\vec \partial} z)^2 = (z{\vec \partial} {\bf v} - {\bf
v}{\vec \partial} z)^2 , \label{eomzv3} \eea
where $\Delta \equiv \partial^2/\partial u_1^2+\partial^2/\partial
u_2^2$ is the world-sheet Laplacian. And the Virasoro constraints
take the following form
\bea & & (\partial_1 z)^2 - (\partial_2 z)^2 + (z \partial_1 {\bf v}
-{\bf v} \partial_1 z)^2 - (z \partial_2 {\bf v} -{\bf v} \partial_2
z)^2 = 0 , \label{vczv1} \\ & & (\partial_1 z) (\partial_2 z) + (z
\partial_1 {\bf v} -{\bf v} \partial_1 z) (z \partial_2 {\bf v}
-{\bf v} \partial_2 z) = 0 . \label{vczv2} \eea
%


For $L = const$, an ansatz of the solution is given in \cite{MMT} as
\bea z = \sum_{a=1}^n z_a e^{\vec k_a \cdot \vec u}, \quad {\bf v} =
\sum_{a=1}^n {\bf v}_a e^{\vec k_a \cdot \vec u}, \label{expzv} \eea
where $n$ is the number of external gluons. And the boundary
conditions are given as
\bea \Delta_a {\bf y} = \frac{{\bf v}_{a+1}}{z_{a+1}} - \frac{{\bf
v}_a}{z_a} = {\bf p}_a , \label{bc} \eea
where ${\bf p}_a$ are the $n$ external momenta.

It is easy to see that the ansatz eq.(\ref{expzv}) satisfies
eq.(\ref{eomzv}) if ${\vec k}_a^2 = L$. Nontrivial equations are
eqs.(\ref{eomzv3})-(\ref{vczv2}). By substitution of
eq.(\ref{expzv}), eq.(\ref{eomzv3}) takes the following form
\bea \sum_{a,b} z_a z_b \Big(L - (\vec k_a \cdot \vec k_b)\Big)
E_{a+b} - \sum_{a<b,~c<d} ({\bf{\cal P}}_{ab}{\bf{\cal
P}}_{cd})(\vec k_{ab} \cdot \vec k_{cd}) E_{a+b+c+d} = 0 ,
\label{eom3} \eea
and the Virasoro constraints eqs.(\ref{vczv1})-(\ref{vczv2}) take
the following form
\bea & & \sum_{a,b} z_a z_b (k_a^1 k_b^1-k_a^2 k_b^2) E_{a+b} +
\sum_{a<b,~c<d} ({\bf{\cal P}}_{ab}{\bf{\cal P}}_{cd})(k_{ab}^1
k_{cd}^1-k_{ab}^2 k_{cd}^2) E_{a+b+c+d} = 0 , \label{vc1} \\ & &
\sum_{a,b} z_a z_b (k_a^1 k_b^2+k_a^2 k_b^1) E_{a+b}
+\sum_{a<b,~c<d} ({\bf{\cal P}}_{ab}{\bf{\cal P}}_{cd})(k_{ab}^1
k_{cd}^2+k_{ab}^2 k_{cd}^1) E_{a+b+c+d} = 0, \label{vc2} \eea
where all the summations are from $1$ to $n$, and
\bea & & E_{a_1+\ldots+a_m} = e^{(\vec k_{a_1}+\ldots+ \vec k_{a_m})
\cdot \vec u}, \quad \vec k_{ab} = \vec k_a-\vec k_b, \quad
\vec k=(k^1, k^2), \nonumber \\
& & {\bf{\cal P}}_{ab} = z_a{\bf v}_b - z_b{\bf v}_a =
z_az_b\big({\bf p}_a+{\bf p}_{a+1} + \ldots + {\bf p}_{b-1}\big) .
\label{calP} \eea
%

\bigskip

Now we try to solve these equations. We first consider
eq.(\ref{eom3}). This equation consists of a summation of a series
of independent $E$-functions ($E_{a+b+...}$). So solving this
equation is equivalent to requiring the vanishing of the coefficient
of each independent $E$-function. Let's first study the terms of
$E_{2a+(a-1)+(a+1)}$. Eq.(\ref{eom3}) requires that
\bea 0 & = & \left({\bf{\cal P}}_{(a-1)a}{\bf{\cal
P}}_{a(a+1)}\right) \left(\vec k_{(a-1)a} \cdot \vec
k_{a(a+1)}\right) E_{2a+(a-1)+(a+1)} \nonumber \\ & = & z_a^2
z_{a-1} z_{a+1} (2{\bf p}_{a-1}{\bf p}_a) \left(\vec k_{(a-1)a}
\cdot \vec k_{a(a+1)}\right) E_{2a+(a-1)+(a+1)} , \quad {\rm for ~~
all}~~ a = 1,2,\dots,n. \eea
where $n+1=1$ by cyclicity. Since $2{\bf p}_{a-1}{\bf p}_a = ({\bf
p}_{a-1}+{\bf p}_a)^2 \neq 0$, the above equations are equivalent to
\bea \vec k_{(a-1)a} \cdot \vec k_{a(a+1)} = (\vec k_{a-1} - \vec
k_a) \cdot (\vec k_a - \vec k_{a+1}) = 0 , \quad {\rm for ~~ all}~~
a = 1,2,\dots,n. \eea
For $\vec k_a^2=L=const$, these conditions can be all satisfied only
at $n=4$, where the four $\vec k$-vectors point along the diagonals
of a rectangle. This indicates that the ansatz
eq.(\ref{expzv}) can \emph{not} be applied to $n>4$ cases directly%
\footnote{We can also study the interesting 3-point case. The
scattering amplitude of three on-shell gluons is identically zero.
In Alday and Maldacena's proposal, this can be understood by
noticing that three lightlike lines can not constitute a triangle.
But if one gluon is off-shell, it is possible to find a solution for
eq.(\ref{eom3}) under the ansatz eq.(\ref{expzv}). However, this
solution has a non-analytic point, and even worse, it is
incompatible with one of the Virasoro constraints.}.
We will only consider the 4-point case.

We can take the ${\vec k}$-vectors generally as
\bea \vec k_1 &=& \sqrt L
(\cos{\phi_1},\sin{\phi_1}),~~~~~~~~~~~~~~~~~~~\vec k_2 =\sqrt L
(\cos{\phi_2},-\sin{\phi_2}),\nonumber\\ \vec k_3 &=& \sqrt L
(-\cos{\phi_1},-\sin{\phi_1}) = -\vec k_{1},~~~~\vec k_4 = \sqrt L
(-\cos{\phi_2},\sin{\phi_2}) = -\vec k_{2}. \label{k} \eea
The parameter $L$ is inessential, due to the scaling
reparametrization invariance of $(u_1,u_2)$. On the other hand, the
parameters $\{\phi_1,\phi_2\}$ are important, since different values
of $\{\phi_1,\phi_2\}$ can correspond to physically inequivalent
solutions.

After substitution of eq.(\ref{k}) for ${\vec k}$-vectors,
eq.(\ref{eom3}) can be collected as
\bea 0 & = & (1-z_1 z_3 s-z_2 z_4 t)
\left[\sin^2\left({\phi_1+\phi_2\over2}\right)(z_1 z_2 E_{1+2}+ z_3
z_4 E_{3+4}) \right.\nonumber
\\ && \left.  + \cos^2\left({\phi_1+\phi_2\over2}\right)
(z_1 z_4 E_{1+4}+ z_2 z_3 E_{2+3}) + (z_1 z_3 + z_2 z_4) E_0 \right]
, \label{eom3collect} \eea
where $s=(p_1+p_3)^2,~ t=(p_2+p_3)^2$ are the Mandelstam variables.
The coefficients of the remaining five independent $E$-functions
have a common factor, so the equation can be (and only be) solved by
requiring this factor to vanish, i.e.
\bea z_1 z_3 s+z_2 z_4 t=1 . \label{zst1} \eea
From this relation, we see that there is still much freedom of
choosing $z_a$, while $\{\phi_1,\phi_2\}$ are totally unfixed.

\bigskip

Next we consider the Virasoro constraints eq.(\ref{vc1}) and
eq.(\ref{vc2}). Similar to solving eq.(\ref{eom3}), we substitute
eq.(\ref{k}) for ${\vec k}$-vectors, and collect the terms for
independent $E$-functions. Then eq.(\ref{vc1}) gives
\bea
&&\left\{\cos(2\phi_1)z_1z_3[1-2z_1z_3s-(s+t)z_2z_4]+\cos(2\phi_2)
z_2z_4[1-2z_2z_4t-(s+t)z_1z_3]\right\}2E_0\nonumber\\&&
-\left[\cos(2\phi_1)z_1z_3s+\cos(2\phi_2)z_2z_4t+\cos(\phi_1-\phi_2)
(1-z_1z_3s-z_2z_4t)\right]2(z_1z_2E_{1+2}+z_3z_4E_{3+4}) \nonumber\\
&&
-\left[\cos(2\phi_1)z_1z_3s+\cos(2\phi_2)z_2z_4t-\cos(\phi_1-\phi_2)
(1-z_1z_3s-z_2z_4t)\right]2(z_1z_4E_{1+4}+z_2z_3E_{2+3}) \nonumber\\
&& -\{\cos(2\phi_1)-[\cos(2\phi_1)-\cos(2\phi_2)] z_2 z_4 t\}(z_1^2
E_{1+1}+ z_3^2 E_{3+3}) \nonumber\\
&&-\{\cos(2\phi_2)+[\cos(2\phi_1)-\cos(2\phi_2)] z_1 z_3 s\}(z_2^2
E_{2+2}+ z_4^2 E_{4+4})= 0 , \label{vcnew1} \eea
and eq.(\ref{vc2}) gives
\bea &&
\left\{\sin(2\phi_1)z_1z_3[1-2z_1z_3s-(s+t)z_2z_4]-\sin(2\phi_2)
z_2z_4[1-2z_2z_4t-(s+t)z_1z_3]\right\}2E_0\nonumber\\&&
-\left[\sin(2\phi_1)z_1z_3s-\sin(2\phi_2)z_2z_4t+\sin(\phi_1-\phi_2)
(1-z_1z_3s-z_2z_4t)\right]2(z_1z_2E_{1+2}+z_3z_4E_{3+4}) \nonumber\\
&&
-\left[\sin(2\phi_1)z_1z_3s-\sin(2\phi_2)z_2z_4t-\sin(\phi_1-\phi_2)
(1-z_1z_3s-z_2z_4t)\right]2(z_1z_4E_{1+4}+z_2z_3E_{2+3})
\nonumber\\&& -\{\sin(2\phi_1)-[\sin(2\phi_1)+\sin(2\phi_2)] z_2 z_4
t\}(z_1^2 E_{1+1}+ z_3^2 E_{3+3}) \nonumber\\ &&
+\{\sin(2\phi_2)-[\sin(2\phi_1)+\sin(2\phi_2)] z_1 z_3 s\}(z_2^2
E_{2+2}+ z_4^2 E_{4+4})=0 . \label{vcnew2} \eea

We first consider the terms of $E_{a+a}$ in the above two equations.
The vanishing of their coefficients is equivalent to the following
relations
\bea & & z_1 z_3 s = {-\cos(2\phi_2)\over
\cos(2\phi_1)-\cos(2\phi_2)} =
{\sin(2\phi_2)\over \sin(2\phi_1)+\sin(2\phi_2)} ~, \nonumber\\
& & z_2 z_4 t = {\cos(2\phi_1)\over \cos(2\phi_1)-\cos(2\phi_2)}=
{\sin(2\phi_1)\over \sin(2\phi_1)+\sin(2\phi_2)} ~, \label{eq1}\eea
from which we get a relation for $\{\phi_1,\phi_2\}$,
\bea 0 = \sin(2\phi_1) \cos(2\phi_2)+ \cos(2\phi_1)
\sin(2\phi_2)=\sin[2(\phi_1+\phi_2)] . \eea
Physically we require that $\phi_1+\phi_2 \neq 0,\pi$, so the above
equation is solved by
\bea \phi_1+\phi_2={\pi\over2} \label{phi} ~. \eea
This also gives that $\sin(2\phi_1)=\sin(2\phi_2)$ and
$\cos(2\phi_1)=-\cos(2\phi_2)$. Substituting this back into
eq.(\ref{eq1}), we get another relation%
\footnote{In the ansatz eq.(\ref{expzv}) of the solution, we don't
require $s,t>0$, which corresponds to spacelike momentum transfer.
However, the constraint eq.(\ref{zst2}) infers it has to be so.
Otherwise, if $z_1 z_3<0$ or $z_2 z_4<0$, there will be singular
points on the boundary for the solutions, which is physically
inconsistent.}
\bea z_1 z_3 s=z_2 z_4 t={1\over 2} ~, \label{zst2} \eea
which also solves eq.(\ref{zst1}) that we have got from solving the
equations of motion.

By using eq.(\ref{phi}) and eq.(\ref{zst2}), we find that all the
coefficients of other $E$-functions in eq.(\ref{vcnew1}) and
eq.(\ref{vcnew2}) also vanish. Therefore, eq.(\ref{phi}) and
eq.(\ref{zst2}) are our final constraints on the solution.

It may be a little surprising that all the three complicated
equations eq.(\ref{eom3collect}), eq.(\ref{vcnew1}) and
eq.(\ref{vcnew2}) lead to only two relations. Under the special
ansatz eq.(\ref{expzv}), we actually have more equations (every
independent $E$-function gives an equation) than the freedom of the
solution, i.e. the solution is overdetermined by the equations of
motion and the Virasoro constraints. Generally, there would be no
solution under this ansatz, such as for  $n>4$ cases.

\bigskip

It seems that we still have freedom to choose the value of $\phi_1$
(or $\phi_2$) in eq.(\ref{phi}), and also have freedom to choose the
value of $z_1$ (or $z_3$) and $z_2$ (or $z_4$) in eq.(\ref{zst2}).
However, all these solutions are equivalent to each other due to two
kinds of reparametrization invariance of $(u_1,u_2)$. By rotational
reparametrization invariance, it's easy to see that only the sum of
$\phi_1$ and $\phi_2$ is physically important; while the
translational reparametrization invariance tells us that the freedom
in $z_a$ is trivial, which we will show explicitly below. Besides an
inessential shift, the vectors ${\bf v}_a$ can also be fixed by the
boundary conditions eq.(\ref{bc}) with given $z_a$. So the solution
is actually unique.

Let's give the explicit form of the solution for $r$. By using
eq.(\ref{phi}) and eq.(\ref{zst2}), we can write $z$ generally as
\bea z=z_1 e^{\vec k_1 \cdot \vec u} + {1\over 2 s z_1} e^{-\vec k_1
\cdot \vec u} + z_2 e^{\vec k_2 \cdot \vec u} + {1\over 2 t z_2}
e^{-\vec k_2 \cdot \vec u} . \label{gen-z} \eea
By choosing a constant world-sheet vector $\vec\delta$ which
satisfies
\bea e^{\vec k_1 \cdot \vec \delta} = \sqrt{2s}~z_1~, \quad e^{\vec
k_2 \cdot \vec \delta} = \sqrt{2t}~z_2 ~, \eea
we can make a translational transformation: $\vec {u'}=\vec u + \vec
\delta$. Then eq.(\ref{gen-z}) reads
\bea z={1\over\sqrt{2s}}\left( e^{\vec k_1 \cdot \vec {u'}} +
e^{-\vec k_1 \cdot \vec {u'}}\right) + {1\over\sqrt{2t}}
\left(e^{\vec k_2 \cdot \vec {u'}} + e^{-\vec k_2 \cdot \vec {u'}}
\right) . \eea
Furthermore, we can set $\phi_1=\phi_2=\pi/4$ by making a rotational
transformation of $(u_1,u_2)$, and also we can set $L=2$ by making a
scaling transformation of $(u_1,u_2)$. Then we have $k_1=(+1,+1)$
and $k_2=(+1,-1)$. After these transformations, we can write $z$ as
\bea z=\left({1\over \sqrt{2s}}+{1\over \sqrt{2t}}\right){2 \cosh
u'_1 \cosh u'_2} + \left({1\over \sqrt{2s}} - {1\over
\sqrt{2t}}\right) {2 \sinh u'_1 \sinh u'_2} ~. \eea
The unique solution for $r$ is then
\bea r={1\over z}={a \over \cosh u'_1 \cosh u'_2 + b \sinh u'_1
\sinh u'_2} ~, \eea
where
\bea a={\sqrt{st}\over \sqrt{2s}+\sqrt{2t}}~, \quad b={\sqrt t
-\sqrt s \over \sqrt t +\sqrt s} ~. \eea
This is exactly the same solution as the one found by Alday and
Maldacena in \cite{AM}.

In \cite{MMT}, the authors found a moduli space $\{z_a, \phi\}$ for
the solutions without considering the Virasoro constraints. By
imposing the Virasoro constraints, we find that there are other
independent relations and the solution can be fixed uniquely. This
also indicates that for a general two-dimensional $\sigma$-model,
where there are no Virasoro constraints, it is possible to have a
larger space of solutions.

\section*{Acknowledgements}
The author would like to thank Wei He and Jun-Bao Wu for
interesting discussions. He would also like to thank Chuan-Jie Zhu
for guidance, discussions and careful reading of the paper. This
work is supported by funds from the National Natural Science
Foundation of China with grant Nos. 10475104 and 10525522.


\end{document}